\documentclass[%
 reprint,
 amsmath,amssymb,
 aps,
]{revtex4-2}

\usepackage{graphicx}
\usepackage{dcolumn}
\usepackage{bm}

\begin{document}
\title{Charge Asymmetry Suppresses Coarsening Dynamics in Polyelectrolyte Complex Coacervation}
\author{Shensheng Chen}
\author{Zhen-Gang Wang}%
 \email{zgw@caltech.edu}
\affiliation{Division of Chemistry and Chemical Engineering, California Institute of Technology, Pasadena, CA 91125}%

\begin{abstract}
Mixing solutions of oppositely charged macromolecules can result in liquid–liquid phase separation into a polymer-rich coacervate phase and a polymer-poor supernatant phase.  Here we show that charge asymmetry in the constituent polymers can slow down the coarsening dynamics, with an apparent growth exponent that deviates from the well-known 1/3 for neutral systems and decreases with increasing degrees of charge asymmetry. Decreasing solvent quality accelerates the coarsening dynamics for asymmetric mixtures, but slows down the coarsening dynamics for symmetric mixtures. We rationalize these results by examining the interaction potential between merging droplets.
\end{abstract}


\maketitle

The association of macromolecules in solution often results in liquid–liquid phase separation (LLPS), with the formation of a polymer-rich liquid phase (coacervate) coexisting with a dilute (supernatant) phase. The liquid-like coacervates, which can be formed by complexation of oppositely charged polyelectrolytes (PE) and or by protein association, are of great importance in materials and biology \cite{Banani2017BiomolecularBiochemistry,Shin2017LiquidDisease}. The formation of these liquid coacervates involves the coarsening of small droplets into larger droplets \cite{Banani2017BiomolecularBiochemistry,Lee2021ChromatinCondensates,Chen2022DrivingCoacervation}. For biocondensates and PE complex coacervates, the coarsening process is experimentally found to be significantly slower than the theoretically anticipated spinodal decomposition dynamics for solutions of uncharged polymers, for which the evolution of domain coarsening follows the well-known power law $R \sim t^{\beta}$ with $\beta=1/3$ \cite{Siggia1979LateMixtures}. For example, Lee, Wingreen and Brangwynne \cite{Lee2021ChromatinCondensates} found the coarsening dynamics in LLPS of biocondensates in the cellular environment is considerably suppressed, with a growth exponent $\beta \approx 0.12 $. The coarsening in biocondensate coacervate usually stops with multiple stable droplets \cite{Shin2018LiquidGenome,Lee2021ChromatinCondensates,Zhang2021MechanicalChromatin}. Liu et al. \cite{Liu2016EarlyScattering,Liu2017StructureSystem} found that the coarsening dynamics is also suppressed in LLPS of PE complex coacervation from mixing oppositely charged PE solutions, with $\beta=0.093 \sim 0.18$.  

The suppressed coarsening dynamics in coacervate formation suggests that the merging of small droplets does not follow the downward free energy path (constrained by material diffusion) as in the classical spinodal decomposition mechanism. Possible explanations to origin of the suppressed dynamics in the LLPS in biocondensate coacervates include the coupling of the phase separation to the elastic deformation of an underlying chromatin network \cite{Lee2021ChromatinCondensates,Zhang2021MechanicalChromatin,Qi2021ChromatinCoalescence} or non-equilibrium activities in the cellular environments \cite{Falahati2017IndependentVivo,Wurtz2018Chemical-Reaction-ControlledCoarsening}. While these mechanisms are reasonable for LLPS dynamics in biocondensates in the cells, they cannot explain the observed deviations from simple spinodal decomposition dynamics in PE complex coacervation. A common feature in both biocondensate coacervates and PE complex coacervates is that these coacervate droplets involve charged macromolecules. The macromolecular charges in the droplets are in general not balanced\cite{Pak2016SequenceProtein,Yin2016Non-equilibriumExcitation,Agrawal2022ManipulationField,Chen2022ComplexationAsymmetry,Crabtree2021RepulsiveCondensates} --  perfect charge balance is the exception rather than the rule. The charge asymmetry between polycations and polyanions (or in the case of biocondensates, the unequal number of positively and negatively charged residues on the biomacromolecules) in general gives rise to net-charged clusters carrying the sign of the overall net charge of the macromolecules\cite{Chen2022ComplexationAsymmetry}. The interaction between charged droplets is expected to create a free energy barrier that can potentially slow down the coarsening dynamics, and in the case of highly asymmetric systems, even stop the coarsening process, resulting in finite-sized clusters \cite{Zhang2005PhasePolyelectrolytes,Chen2022ComplexationAsymmetry}.

In this letter, we examine the effects of charge asymmetry on the dynamics of LLPS by simulating the coarsening of PE coacervate droplets under different charge asymmetry and solvent quality conditions using electrostatic dissipative particle dynamics (EDPD) simulation \cite{Groot2003ElectrostaticSurfactants,Chen2022ComplexationAsymmetry}. In our DPD simulation, monomers of polycations and polyanions, their counterions, and solvent (water) are explicitly modeled by coarse-grained beads having the same mass $m$ and size $r_c$. Non-bonded beads interact with each other via a soft repulsive force given by $\bm{F}_{ij}^{C}=a_{ij}(1-r_{ij}/r_{c})\bm{e}_{ij}$, where $a_{ij}$ is the strength of the interaction, $r_{ij}=|\bm{r}_{i}-\bm{r}_{j}|$ is the inter-bead distance, and $\bm{e}_{ij}=(\bm{r}_{i}-\bm{r}_{j})/r_{ij}$ represents the direction of the force. We set the repulsion $a_{ij}$ between all species as $a_{ij}=75 k_BT/r_c$, with the exception of repulsion between the monomer on the polyion and water $a_{pw}$, whose value is given by $a_{pw}=75 k_BT/r_c + \Delta a$ to control the solvent condition by adjusting $\Delta a$. Here, $k_BT$ the thermal energy and we take $T$ to be the room temperature. $\Delta a \le 0$ corresponds to good solvent, and increasing $\Delta a$ decreases the solvent quality.  Although it is possible to map the values of $\Delta a$ to the more familiar Flory--Huggins parameter, such a mapping more appropriate for neutral systems and requires fitting to a particular model.  Here we will simply report the value of $\Delta a$. The thermostat in DPD is controlled by two additional pair-wised forces, the dissipative force and the random force, which are related by the fluctuation-dissipation theorem, with the force parameters given by Groot and Warren \cite{Groot1997DissipativeSimulation}. Since all the forces in the DPD are pair-wise, momentum conservation is guaranteed, allowing hydrodynamics to be captured at the mesoscale. 

The polymers are modeled as beads connected by a harmonic bond potential $E_{bond}=\frac{1}{2} K_{bond}(r-r_{0})^{2}$ between two consecutive beads along linear chains. The bond strength and equilibrium bond length are set to $K_{bond}=100k_{B}T/r_{c}^2$ and $r_0=0.5r_c$, respectively. $m$, $k_BT$, $r_c$ represent the mass, energy and length units in DPD simulations, which are all set to unity. With the typical interpretation that one DPD bead corresponds to three water molecules \cite{Groot1997DissipativeSimulation,Groot2003ElectrostaticSurfactants}, $r_c$ has a physical length of $0.64$nm. Each monomer in polycations/polyanions carries unit charge of $+e$/$-e$, and the corresponding counterion has the opposite unit charge. The Bjerrum length is set to $l_B=0.7$nm to represent the electrostatic interaction strength in water at room temperature. Electrostatic interactions are calculated by solving the Poisson equation in real space in the simulation domain, in which electric field is computed on a  mesh with smeared charges from the off-lattice point charges, as detailed in Ref. \cite{Groot2003ElectrostaticSurfactants}. The total monomer concentration is about 0.14M.  Together with the counterions, the total ion concentration (cation and anion) is 0.28M, corresponding to a nominal Debye screening length of about $1$nm.  Other simulation details are given in Supplementary Material. 

\begin{figure}
\centering
\includegraphics[width=.95\linewidth]{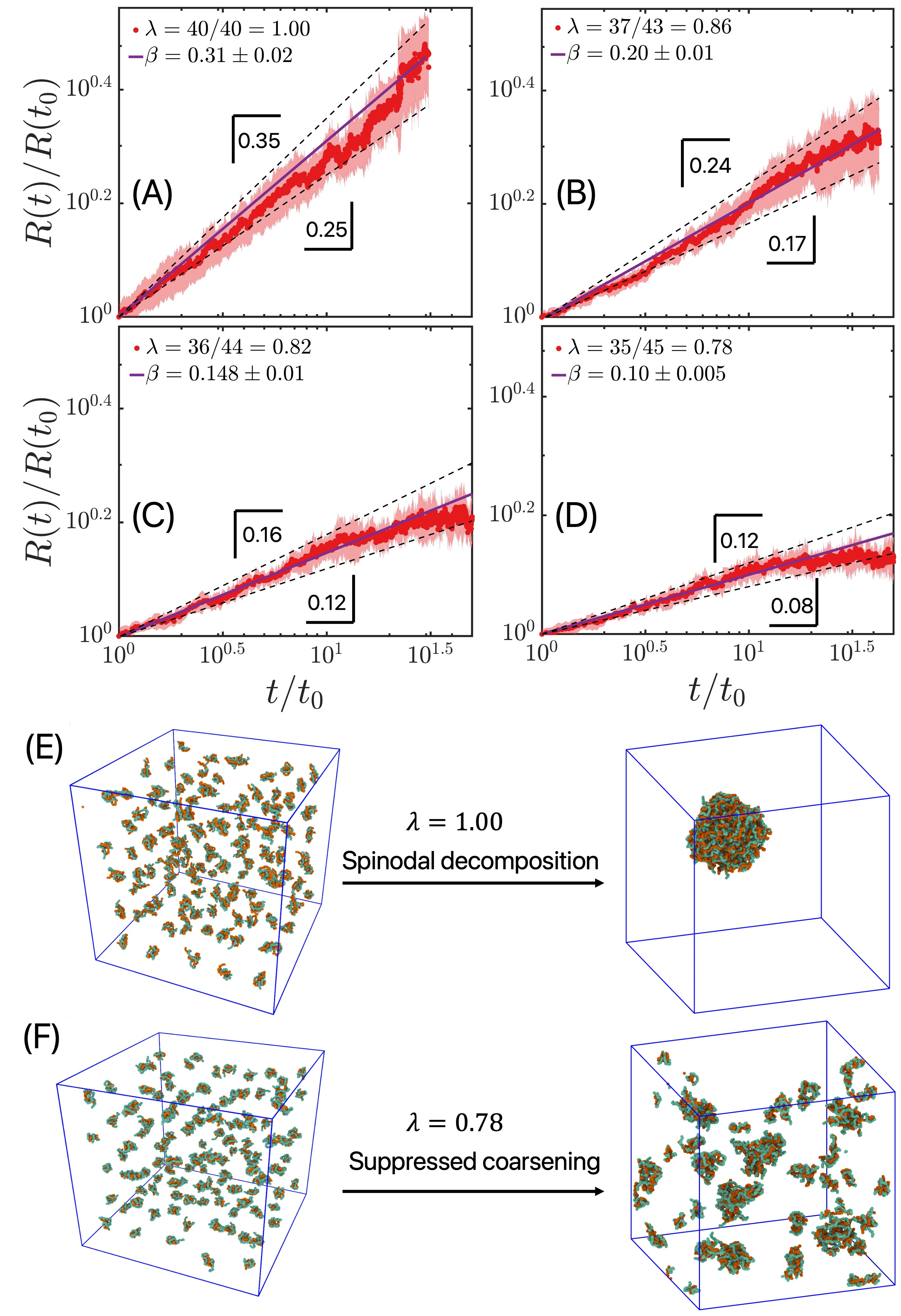}
\caption{A-D: Domain growth in systems with charge ratio (between polycation and polyanion) $\lambda=1.00,0.86,0.82, 0.78$. The solid lines show the best-fitted scaling exponents while the dash lines are the upper and lower bounds of the exponents. $t_0 = 400, 165, 125$, and $100$ in DPD time, from A to E, respectively. The shaded areas represent standard deviation from 10 independent runs. E-F: Simulation snapshots before and after domain coarsening for $\lambda=1.00$ and $\lambda=0.78$, respectively. The green and orange beads represent polycation and polyanion monomers, respectively. Solvent molecules and counterions are not shown for clarity.}
\label{fig:p2}
\end{figure}

Our simulations include 125 fully charged polycations with chain length $N_1$ and 125 polyanions with chain length $N_2$ in a box of $60 \times 60  \times 60 r_c^3$.  The elementary process of polyelectrolyte complex coacevation consists of the formation of polycation--polyanion pairs followed by the coalescence of the pairs into larger clusters \cite{Chen2022DrivingCoacervation,Delaney2017TheorySelf-coacervates}. Since the experiments only monitored the coarsening process, the initial states right after the mixing of the polycation and polyanion solutions is unknown. We thus initialize our system by having 125 small droplets each consisting of a polycation--polyanion pair. Similar to the experimental measurements \cite{Lee2021ChromatinCondensates,Zhang2021MechanicalChromatin}, we monitor the coarsening dynamics by calculating the time evolution of the (number) average radius of gyration $\langle R \rangle$ of  the clusters in the system. The degree of charge asymmetry is defined by $\lambda =N_1/N_2$ where for concreteness we take $N_1 \le N_2$ and keep $N_1+N_2=80$.  For each charge asymmetry, we perform 10 independent runs to obtain the ensemble average.   

We first consider the effects of charge asymmetry $\lambda$ under good solvent condition ($\Delta a = 0$). Figure 1A-1D show the time the evolution of $ R (t) / R(t_0) $ for different values of $\lambda$, where $R(t_0)$ is the average radius of gyration in the systems at time $t_0$ when the size first shows power-law growth. For the charge-balanced system $(\lambda=1)$, Fig. 1A shows the coarsening dynamics follows a power-law scaling $\sim t^{0.31}$, which is very close to the theoretically expected $1/3$ scaling for spinodal decomposition. The slight deviation from $1/3$ is likely due to the finite size effect in simulations (since there are very few droplets in the later stages of the simulation). In all the runs for the charge-balanced systems, the coarsening ends up with the formation of a single large droplet containing all the polyions, as shown in Fig 1E. With charge asymmetry, the phase separation dynamics slows down significantly: The apparent power-law $\sim t^{\beta}$ in domain growth has a smaller exponent of $\beta=0.20, 0.14$, and $0.10$ respectively for $\lambda=0.86,0.82,0.78$. These values  are quite close to the range of the exponents observed in the experiments\cite{Lee2021ChromatinCondensates,Liu2016EarlyScattering}. For $\lambda=0.82$ and $0.78$, the coarsening stops with multiple droplets carrying net negative charges as the final state; a snapshot is shown in Fig. 1F. This result is consistent with experimental observation and with our recent finding on the equilibrium behavior in the complex coacervation of charge-asymmetric polyelectrolyte mixtures \cite{Chen2022ComplexationAsymmetry,Agrawal2022ManipulationField}. For the case of $\lambda=0.86$, the final state consists of 3~5 clusters. On the other hand, if we artificially put all the polymers into a single droplet, the droplet remains stable without spontaneous fission. Whether the final state obtained in our coarsening simulation represents the equilibrium state, or a metastable state due to kinetic barriers for fusion, remains to be investigated further.

To understand the observed slowdown in the coarsening dynamics due to charge asymmetry, we examine the elementary process of the merging of two polyion pairs into a two-pair cluster. To this end, we compute the potential of mean force (PMF) between the two polyion pairs as a function of their center-of-mass distance $r$; the result is shown in Fig. 2, 
\begin{figure}
\centering
\includegraphics[width=.95\linewidth]{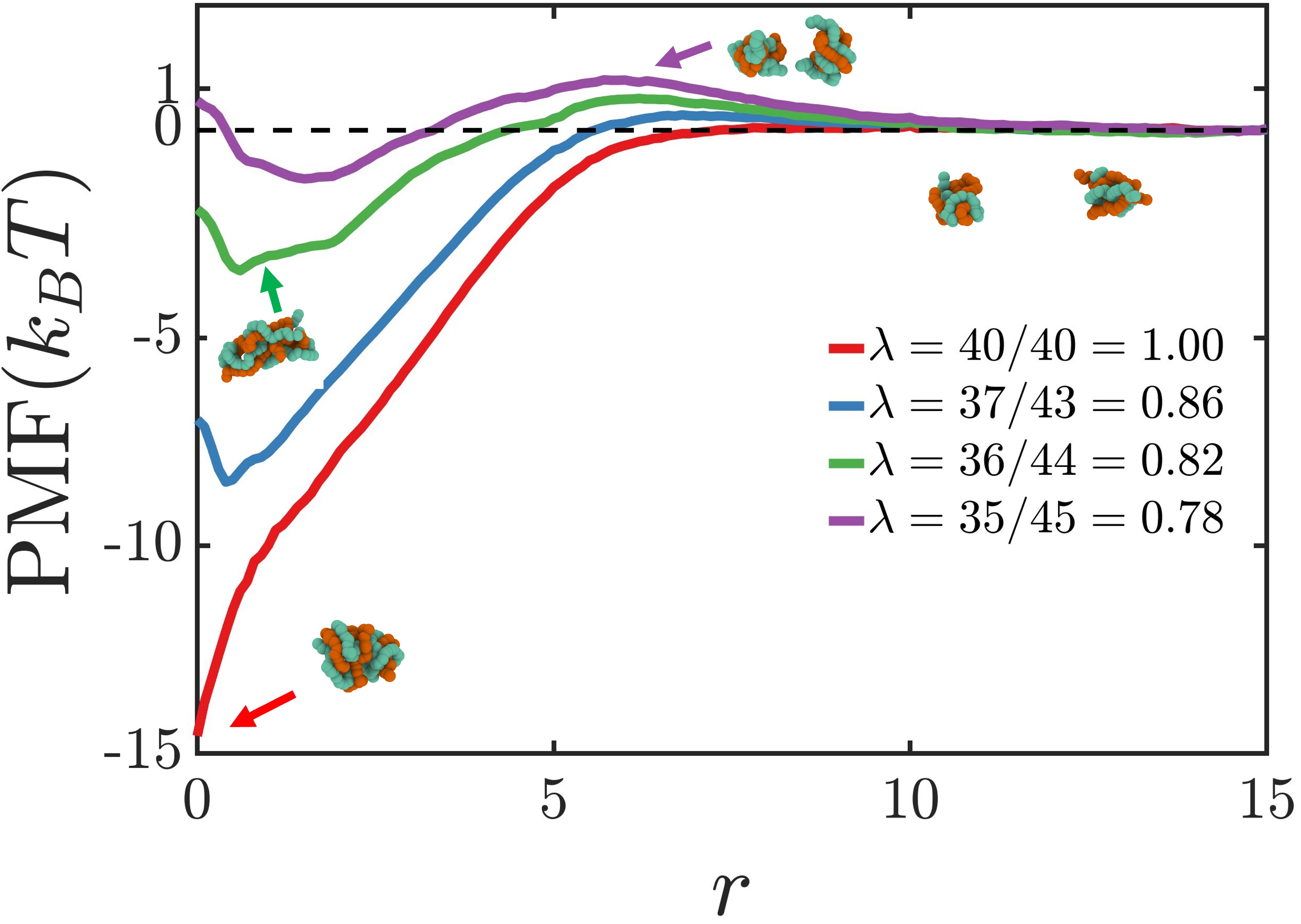}
\caption{Potential of mean force for complexation of two polyion pairs under different charge ratios $\lambda=1.00,0.86,0.82, 0.78$ in good solvent condition. }
\label{fig:p1}
\end{figure}
for the case of good solvent condition ($\Delta a=0$) under different charge ratios. 

For the symmetric system ($\lambda=1$), the coalescence of the two droplets experiences no free energy barrier, consistent with the spinodal decomposition mechanism. For all the three cases with charge asymmetry, coalescence of the two droplets involves an energy barrier along the pathway, with the barrier height increasing with increasing charge asymmetry (i.e., decreasing $\lambda$), consistent with the slower coarsening dynamics for the more charge-asymmetric systems. Clearly, the free energy barrier must arise from the repulsion from droplets carrying net macromolecular charges. Note that the free energy barriers are of order $k_B T$ or less, thus thermal fluctuation can easily drive the fusion of these polyion pairs.  As the droplets grow and accumulate more net charge, we expect the free energy barrier for coalescence to increase, which can eventually arrest further growth. Note that macromolecular condensates carrying net charges can have preferred size as the final equilibrium state or as an intermediate metastable state \cite{Hutchens2007MetastableSolutions}. So the final state observed in our simulation for these asymmetric systems can be due to both thermodynamic and kinetic reasons.   

Theoretically, droplet coalescence and coarsening are driven by the tendency to decrease the surface energy.  Therefore, we expect that coarsening dynamics should be faster with decreasing solvent quality\cite{Siggia1979LateMixtures}. 
\begin{figure}
\centering
\includegraphics[width=.98\linewidth]{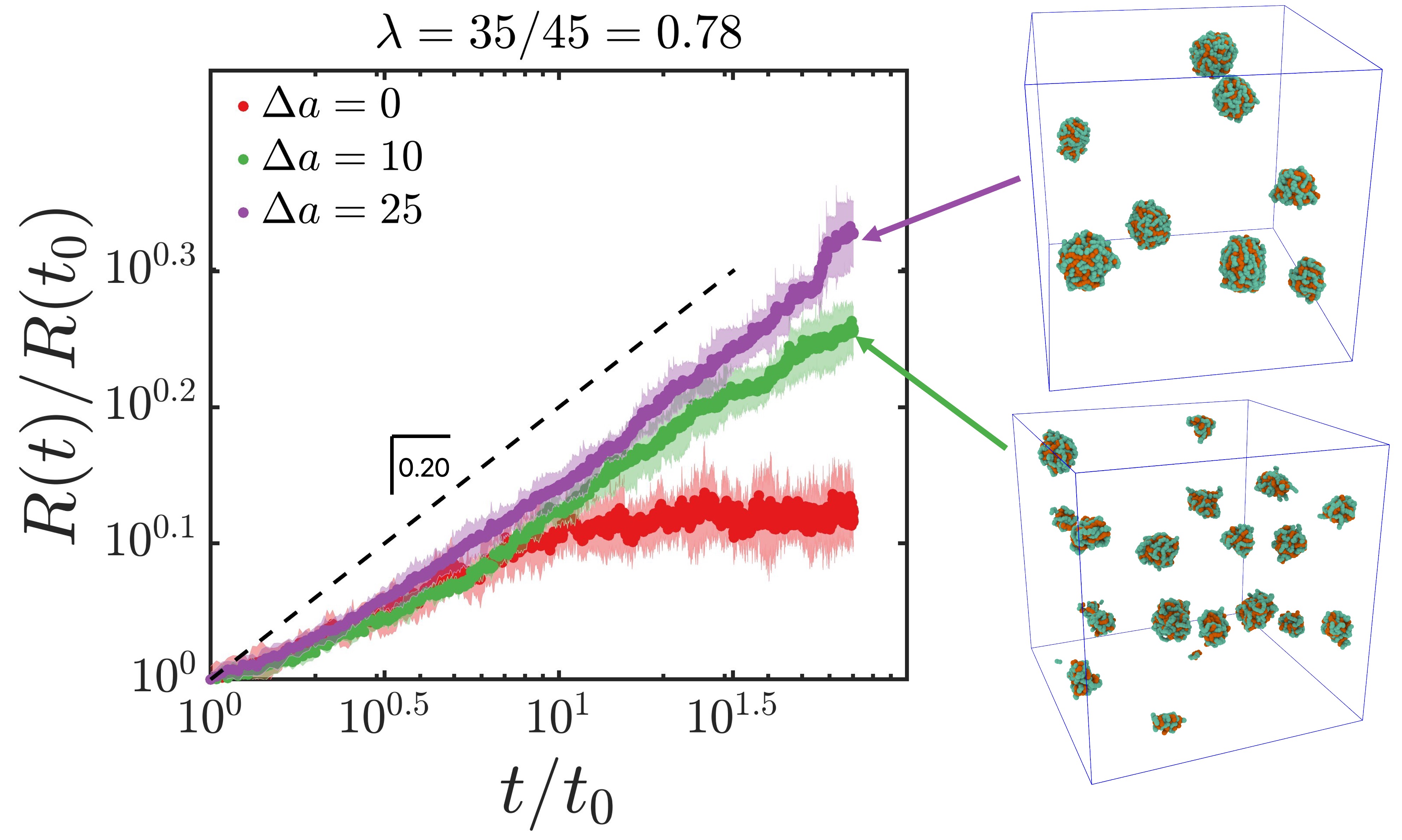}
\caption{Domain growth in systems with charge ratio $\lambda=0.78$ for solvent conditions at $\Delta a =0, 10, 25$, respectively. The simulation snapshots on the right shows the final morphology of the systems containing multiple net-charged droplets.}
\label{fig:p3}
\end{figure}
To this end, we study the LLPS dynamics in systems with charge ratio $\lambda=0.78$ under different solvent conditions with $\Delta a=0, 10, 25$. Figure 3 shows the coarsening dynamics is indeed accelerated -- as evidenced by larger droplet sizes and larger apparent power-law exponent -- with increasing $\Delta a$. In addition, coarsening proceeds further, since the final state contains fewer and larger droplets. However, even for $\Delta a=25$, which corresponds to a very poor solvent, the growth exponent reaches about $\beta=2.0$, significantly below the $\sim 1/3$ observed in charge-balanced systems. These results suggest that while poor solvent conditions can accelerate the coarsening dynamics, electrostatic repulsion between droplets in asymmetric systems still have strong effects in suppressing the coarsening dynamics and preventing the complete coalescence of all the droplets. 



\begin{figure}
\centering
\includegraphics[width=.95\linewidth]{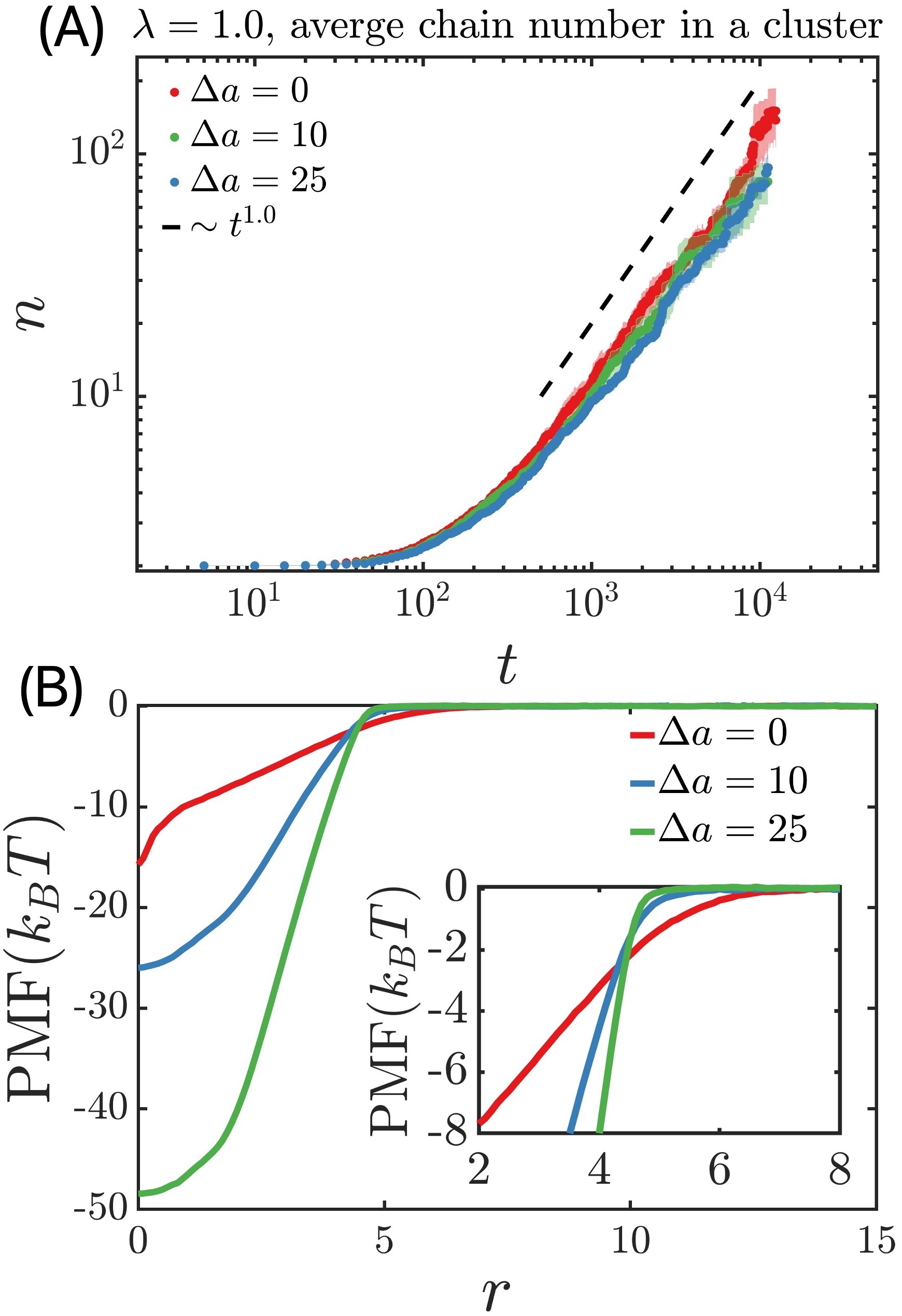}
\caption{A: Time evolution of chain number in a cluster in charge-balanced systems under solvent conditions of  $\Delta a=0,10,25$. The coarsening dynamics in general follow the theoretical $1/3$ domain growth law. With increasing $\Delta a$, the domain growth is slightly slower. B: PMF profile for complexation of two charge-balance pairs under solvent condition of charge asymmetry of  $\Delta a=0,10,25$. Although larger $\Delta a$ gives deeper attractive wells, smaller $\Delta a$ has larger distance to trigger the attraction between the droplets, which can be seen more clearly in the inset.}
\label{fig:p4}
\end{figure}

The effects of solvent condition on LLPS dynamics in the charge-balanced systems are shown in Fig. 4A, where we plot of time evolution of the average number of chains $n$ in a cluster. Not surprisingly, the coarsening dynamics in all the three solvent conditions approximately follow the $t^{1/3}$ domain growth law (noting that $n \sim R^3 \sim t^{3 \beta}$), consistent with the barrierless process for spinodal decomposition; see Fig. 4B. Surprisingly, however, decreasing the solvent quality results in a slight but clearly noticeable slowdown of the coarsening dynamics. This is quite different from the expected behavior for LLPS in neutral systems where decreasing the solvent quality generally result in faster coarsening dynamics \cite{Siggia1979LateMixtures}. 

This anomalous slowdown in coarsening dynamics with decreasing solvent quality can be understood by examining the driving force for the fusion of two polyion pairs. In our recent work, we showed that the mutual polarization between the two polyion globules -- akin to the dispersion force between neutral atoms -- provides a weak attractive force prior to contact\cite{Chen2022DrivingCoacervation}. This polarization interaction is stronger under the good solvent condition compared to the poor solvent conditions due to the smaller surface tension and lower polymer density in the former; see Fig. S1 in Supplementary Material. In addition, the smaller surface tension and lower polymer density make the droplets more susceptible to shape fluctuation, thus facilitating contacts that  trigger the coalescence. We note that these two factors are not independent of each other and it may not be possible to examine them separately. 
The net interaction is obviously the combined effects of the two, as manifested in Fig. 4B, where it can be seen that the attraction between two polyion pairs under good solvent condition sets in at larger distances. 

In summary, we have shown that the charge asymmetry significantly suppresses the coarsening dynamics in polyelectrolyte complex coacervation. Decreasing the solvent quality can accelerate coarsening in charge-asymmetric systems, but cannot recover the $t^{1/3}$ growth law for the classical spinodal decomposition.  For the charge-balanced system, however, decreasing the solvent quality slows down the coarsening dynamics while still maintaining the $t^{1/3}$ power law growth.  While these results are obtained for LLPS dynamics in polyelectrolyte complex coacervation, we expect our findings to be relevant to biocondensates which involve biomacromolecules that are not charge balanced.  Although biocondensate formation in cells  involves other factors such as coupling to the background chromatin network and cellular activities, our study suggests that electrostatics can be a significant contributor in the observed anomalous dynamics. It has been recognized that biocondensate droplets carry surface charges and generate significant electric field in the interfacial region \cite{Welsh2022SurfaceCondensates}, as a result of various sources of charge asymmetry in the cellular environments. Such interfacial electric field has been shown to have biochemical functions, such as in modulating redox reactions \cite{Dai2023InterfaceReactions} and in regulating granule assembly \cite{Hu2021ElectrochemicalGranules}. More studies are warranted to explore the effects of electrostatics in LLPS in cells and in biomimetic systems.


\begin{acknowledgments}
This research is supported by funding from Hong Kong Quantum AI Lab Ltd. We thank the general computation time allocated by the resources of the Center for Functional Nanomaterials (CFN), which is a U.S. Department of Energy Office of Science User Facility, at Brookhaven National Laboratory under Contract No. DE-SC0012704.  We thank Dr. Andrew Ylitalo for useful discussions.
\end{acknowledgments}

\bibliography{spinodal}

\begin{thebibliography}{25}%
\makeatletter
\providecommand \@ifxundefined [1]{%
 \@ifx{#1\undefined}
}%
\providecommand \@ifnum [1]{%
 \ifnum #1\expandafter \@firstoftwo
 \else \expandafter \@secondoftwo
 \fi
}%
\providecommand \@ifx [1]{%
 \ifx #1\expandafter \@firstoftwo
 \else \expandafter \@secondoftwo
 \fi
}%
\providecommand \natexlab [1]{#1}%
\providecommand \enquote  [1]{``#1''}%
\providecommand \bibnamefont  [1]{#1}%
\providecommand \bibfnamefont [1]{#1}%
\providecommand \citenamefont [1]{#1}%
\providecommand \href@noop [0]{\@secondoftwo}%
\providecommand \href [0]{\begingroup \@sanitize@url \@href}%
\providecommand \@href[1]{\@@startlink{#1}\@@href}%
\providecommand \@@href[1]{\endgroup#1\@@endlink}%
\providecommand \@sanitize@url [0]{\catcode `\\12\catcode `\$12\catcode
  `\&12\catcode `\#12\catcode `\^12\catcode `\_12\catcode `\%12\relax}%
\providecommand \@@startlink[1]{}%
\providecommand \@@endlink[0]{}%
\providecommand \url  [0]{\begingroup\@sanitize@url \@url }%
\providecommand \@url [1]{\endgroup\@href {#1}{\urlprefix }}%
\providecommand \urlprefix  [0]{URL }%
\providecommand \Eprint [0]{\href }%
\providecommand \doibase [0]{https://doi.org/}%
\providecommand \selectlanguage [0]{\@gobble}%
\providecommand \bibinfo  [0]{\@secondoftwo}%
\providecommand \bibfield  [0]{\@secondoftwo}%
\providecommand \translation [1]{[#1]}%
\providecommand \BibitemOpen [0]{}%
\providecommand \bibitemStop [0]{}%
\providecommand \bibitemNoStop [0]{.\EOS\space}%
\providecommand \EOS [0]{\spacefactor3000\relax}%
\providecommand \BibitemShut  [1]{\csname bibitem#1\endcsname}%
\let\auto@bib@innerbib\@empty
\bibitem [{\citenamefont {Banani}\ \emph {et~al.}(2017)\citenamefont {Banani},
  \citenamefont {Lee}, \citenamefont {Hyman},\ and\ \citenamefont
  {Rosen}}]{Banani2017BiomolecularBiochemistry}%
  \BibitemOpen
  \bibfield  {author} {\bibinfo {author} {\bibfnamefont {S.~F.}\ \bibnamefont
  {Banani}}, \bibinfo {author} {\bibfnamefont {H.~O.}\ \bibnamefont {Lee}},
  \bibinfo {author} {\bibfnamefont {A.~A.}\ \bibnamefont {Hyman}},\ and\
  \bibinfo {author} {\bibfnamefont {M.~K.}\ \bibnamefont {Rosen}},\ }\bibfield
  {title} {\bibinfo {title} {{Biomolecular condensates: organizers of cellular
  biochemistry}},\ }\href {https://doi.org/10.1038/nrm.2017.7} {\bibfield
  {journal} {\bibinfo  {journal} {Nature Reviews Molecular Cell Biology}\
  }\textbf {\bibinfo {volume} {18}},\ \bibinfo {pages} {285} (\bibinfo {year}
  {2017})}\BibitemShut {NoStop}%
\bibitem [{\citenamefont {Shin}\ and\ \citenamefont
  {Brangwynne}(2017)}]{Shin2017LiquidDisease}%
  \BibitemOpen
  \bibfield  {author} {\bibinfo {author} {\bibfnamefont {Y.}~\bibnamefont
  {Shin}}\ and\ \bibinfo {author} {\bibfnamefont {C.~P.}\ \bibnamefont
  {Brangwynne}},\ }\bibfield  {title} {\bibinfo {title} {{Liquid phase
  condensation in cell physiology and disease}},\ }\href
  {https://doi.org/10.1126/science.aaf4382} {\bibfield  {journal} {\bibinfo
  {journal} {Science}\ }\textbf {\bibinfo {volume} {357}},\ \bibinfo {pages}
  {aaf4382} (\bibinfo {year} {2017})}\BibitemShut {NoStop}%
\bibitem [{\citenamefont {Lee}\ \emph {et~al.}(2021)\citenamefont {Lee},
  \citenamefont {Wingreen},\ and\ \citenamefont
  {Brangwynne}}]{Lee2021ChromatinCondensates}%
  \BibitemOpen
  \bibfield  {author} {\bibinfo {author} {\bibfnamefont {D.~S.~W.}\
  \bibnamefont {Lee}}, \bibinfo {author} {\bibfnamefont {N.~S.}\ \bibnamefont
  {Wingreen}},\ and\ \bibinfo {author} {\bibfnamefont {C.~P.}\ \bibnamefont
  {Brangwynne}},\ }\bibfield  {title} {\bibinfo {title} {{Chromatin mechanics
  dictates subdiffusion and coarsening dynamics of embedded condensates}},\
  }\href {https://doi.org/10.1038/s41567-020-01125-8} {\bibfield  {journal}
  {\bibinfo  {journal} {Nature Physics}\ }\textbf {\bibinfo {volume} {17}},\
  \bibinfo {pages} {531} (\bibinfo {year} {2021})}\BibitemShut {NoStop}%
\bibitem [{\citenamefont {Chen}\ and\ \citenamefont
  {Wang}(2022)}]{Chen2022DrivingCoacervation}%
  \BibitemOpen
  \bibfield  {author} {\bibinfo {author} {\bibfnamefont {S.}~\bibnamefont
  {Chen}}\ and\ \bibinfo {author} {\bibfnamefont {Z.-G.}\ \bibnamefont
  {Wang}},\ }\bibfield  {title} {\bibinfo {title} {{Driving force and pathway
  in polyelectrolyte complex coacervation}},\ }\href
  {https://doi.org/10.1073/pnas.2209975119} {\bibfield  {journal} {\bibinfo
  {journal} {Proceedings of the National Academy of Sciences}\ }\textbf
  {\bibinfo {volume} {119}},\ \bibinfo {pages} {e2209975119} (\bibinfo {year}
  {2022})}\BibitemShut {NoStop}%
\bibitem [{\citenamefont {Siggia}(1979)}]{Siggia1979LateMixtures}%
  \BibitemOpen
  \bibfield  {author} {\bibinfo {author} {\bibfnamefont {E.~D.}\ \bibnamefont
  {Siggia}},\ }\bibfield  {title} {\bibinfo {title} {{Late stages of spinodal
  decomposition in binary mixtures}},\ }\href
  {https://doi.org/10.1103/PhysRevA.20.595} {\bibfield  {journal} {\bibinfo
  {journal} {Physical Review A}\ }\textbf {\bibinfo {volume} {20}},\ \bibinfo
  {pages} {595} (\bibinfo {year} {1979})}\BibitemShut {NoStop}%
\bibitem [{\citenamefont {Shin}\ \emph {et~al.}(2018)\citenamefont {Shin},
  \citenamefont {Chang}, \citenamefont {Lee}, \citenamefont {Berry},
  \citenamefont {Sanders}, \citenamefont {Ronceray}, \citenamefont {Wingreen},
  \citenamefont {Haataja},\ and\ \citenamefont
  {Brangwynne}}]{Shin2018LiquidGenome}%
  \BibitemOpen
  \bibfield  {author} {\bibinfo {author} {\bibfnamefont {Y.}~\bibnamefont
  {Shin}}, \bibinfo {author} {\bibfnamefont {Y.-C.}\ \bibnamefont {Chang}},
  \bibinfo {author} {\bibfnamefont {D.~S.}\ \bibnamefont {Lee}}, \bibinfo
  {author} {\bibfnamefont {J.}~\bibnamefont {Berry}}, \bibinfo {author}
  {\bibfnamefont {D.~W.}\ \bibnamefont {Sanders}}, \bibinfo {author}
  {\bibfnamefont {P.}~\bibnamefont {Ronceray}}, \bibinfo {author}
  {\bibfnamefont {N.~S.}\ \bibnamefont {Wingreen}}, \bibinfo {author}
  {\bibfnamefont {M.}~\bibnamefont {Haataja}},\ and\ \bibinfo {author}
  {\bibfnamefont {C.~P.}\ \bibnamefont {Brangwynne}},\ }\bibfield  {title}
  {\bibinfo {title} {{Liquid Nuclear Condensates Mechanically Sense and
  Restructure the Genome}},\ }\href
  {https://doi.org/10.1016/j.cell.2018.10.057} {\bibfield  {journal} {\bibinfo
  {journal} {Cell}\ }\textbf {\bibinfo {volume} {175}},\ \bibinfo {pages}
  {1481} (\bibinfo {year} {2018})}\BibitemShut {NoStop}%
\bibitem [{\citenamefont {Zhang}\ \emph {et~al.}(2021)\citenamefont {Zhang},
  \citenamefont {Lee}, \citenamefont {Meir}, \citenamefont {Brangwynne},\ and\
  \citenamefont {Wingreen}}]{Zhang2021MechanicalChromatin}%
  \BibitemOpen
  \bibfield  {author} {\bibinfo {author} {\bibfnamefont {Y.}~\bibnamefont
  {Zhang}}, \bibinfo {author} {\bibfnamefont {D.~S.}\ \bibnamefont {Lee}},
  \bibinfo {author} {\bibfnamefont {Y.}~\bibnamefont {Meir}}, \bibinfo {author}
  {\bibfnamefont {C.~P.}\ \bibnamefont {Brangwynne}},\ and\ \bibinfo {author}
  {\bibfnamefont {N.~S.}\ \bibnamefont {Wingreen}},\ }\bibfield  {title}
  {\bibinfo {title} {{Mechanical Frustration of Phase Separation in the Cell
  Nucleus by Chromatin}},\ }\href
  {https://doi.org/10.1103/PhysRevLett.126.258102} {\bibfield  {journal}
  {\bibinfo  {journal} {Physical Review Letters}\ }\textbf {\bibinfo {volume}
  {126}},\ \bibinfo {pages} {258102} (\bibinfo {year} {2021})}\BibitemShut
  {NoStop}%
\bibitem [{\citenamefont {Liu}\ \emph {et~al.}(2016)\citenamefont {Liu},
  \citenamefont {Haddou}, \citenamefont {Grillo}, \citenamefont {Mana},
  \citenamefont {Chapel},\ and\ \citenamefont
  {Schatz}}]{Liu2016EarlyScattering}%
  \BibitemOpen
  \bibfield  {author} {\bibinfo {author} {\bibfnamefont {X.}~\bibnamefont
  {Liu}}, \bibinfo {author} {\bibfnamefont {M.}~\bibnamefont {Haddou}},
  \bibinfo {author} {\bibfnamefont {I.}~\bibnamefont {Grillo}}, \bibinfo
  {author} {\bibfnamefont {Z.}~\bibnamefont {Mana}}, \bibinfo {author}
  {\bibfnamefont {J.-P.}\ \bibnamefont {Chapel}},\ and\ \bibinfo {author}
  {\bibfnamefont {C.}~\bibnamefont {Schatz}},\ }\bibfield  {title} {\bibinfo
  {title} {{Early stage kinetics of polyelectrolyte complex coacervation
  monitored through stopped-flow light scattering}},\ }\href
  {https://doi.org/10.1039/C6SM01979J} {\bibfield  {journal} {\bibinfo
  {journal} {Soft Matter}\ }\textbf {\bibinfo {volume} {12}},\ \bibinfo {pages}
  {9030} (\bibinfo {year} {2016})}\BibitemShut {NoStop}%
\bibitem [{\citenamefont {Liu}\ \emph {et~al.}(2017)\citenamefont {Liu},
  \citenamefont {Chapel},\ and\ \citenamefont
  {Schatz}}]{Liu2017StructureSystem}%
  \BibitemOpen
  \bibfield  {author} {\bibinfo {author} {\bibfnamefont {X.}~\bibnamefont
  {Liu}}, \bibinfo {author} {\bibfnamefont {J.-P.}\ \bibnamefont {Chapel}},\
  and\ \bibinfo {author} {\bibfnamefont {C.}~\bibnamefont {Schatz}},\
  }\bibfield  {title} {\bibinfo {title} {{Structure, thermodynamic and kinetic
  signatures of a synthetic polyelectrolyte coacervating system}},\ }\href
  {https://doi.org/10.1016/j.cis.2016.10.004} {\bibfield  {journal} {\bibinfo
  {journal} {Advances in Colloid and Interface Science}\ }\textbf {\bibinfo
  {volume} {239}},\ \bibinfo {pages} {178} (\bibinfo {year}
  {2017})}\BibitemShut {NoStop}%
\bibitem [{\citenamefont {Qi}\ and\ \citenamefont
  {Zhang}(2021)}]{Qi2021ChromatinCoalescence}%
  \BibitemOpen
  \bibfield  {author} {\bibinfo {author} {\bibfnamefont {Y.}~\bibnamefont
  {Qi}}\ and\ \bibinfo {author} {\bibfnamefont {B.}~\bibnamefont {Zhang}},\
  }\bibfield  {title} {\bibinfo {title} {{Chromatin network retards nucleoli
  coalescence}},\ }\href {https://doi.org/10.1038/s41467-021-27123-9}
  {\bibfield  {journal} {\bibinfo  {journal} {Nature Communications}\ }\textbf
  {\bibinfo {volume} {12}},\ \bibinfo {pages} {6824} (\bibinfo {year}
  {2021})}\BibitemShut {NoStop}%
\bibitem [{\citenamefont {Falahati}\ and\ \citenamefont
  {Wieschaus}(2017)}]{Falahati2017IndependentVivo}%
  \BibitemOpen
  \bibfield  {author} {\bibinfo {author} {\bibfnamefont {H.}~\bibnamefont
  {Falahati}}\ and\ \bibinfo {author} {\bibfnamefont {E.}~\bibnamefont
  {Wieschaus}},\ }\bibfield  {title} {\bibinfo {title} {{Independent active and
  thermodynamic processes govern the nucleolus assembly in vivo}},\ }\href
  {https://doi.org/10.1073/pnas.1615395114} {\bibfield  {journal} {\bibinfo
  {journal} {Proceedings of the National Academy of Sciences}\ }\textbf
  {\bibinfo {volume} {114}},\ \bibinfo {pages} {1335} (\bibinfo {year}
  {2017})}\BibitemShut {NoStop}%
\bibitem [{\citenamefont {Wurtz}\ and\ \citenamefont
  {Lee}(2018)}]{Wurtz2018Chemical-Reaction-ControlledCoarsening}%
  \BibitemOpen
  \bibfield  {author} {\bibinfo {author} {\bibfnamefont {J.~D.}\ \bibnamefont
  {Wurtz}}\ and\ \bibinfo {author} {\bibfnamefont {C.~F.}\ \bibnamefont
  {Lee}},\ }\bibfield  {title} {\bibinfo {title} {{Chemical-Reaction-Controlled
  Phase Separated Drops: Formation, Size Selection, and Coarsening}},\ }\href
  {https://doi.org/10.1103/PhysRevLett.120.078102} {\bibfield  {journal}
  {\bibinfo  {journal} {Physical Review Letters}\ }\textbf {\bibinfo {volume}
  {120}},\ \bibinfo {pages} {078102} (\bibinfo {year} {2018})}\BibitemShut
  {NoStop}%
\bibitem [{\citenamefont {Pak}\ \emph {et~al.}(2016)\citenamefont {Pak},
  \citenamefont {Kosno}, \citenamefont {Holehouse}, \citenamefont {Padrick},
  \citenamefont {Mittal}, \citenamefont {Ali}, \citenamefont {Yunus},
  \citenamefont {Liu}, \citenamefont {Pappu},\ and\ \citenamefont
  {Rosen}}]{Pak2016SequenceProtein}%
  \BibitemOpen
  \bibfield  {author} {\bibinfo {author} {\bibfnamefont {C.}~\bibnamefont
  {Pak}}, \bibinfo {author} {\bibfnamefont {M.}~\bibnamefont {Kosno}}, \bibinfo
  {author} {\bibfnamefont {A.}~\bibnamefont {Holehouse}}, \bibinfo {author}
  {\bibfnamefont {S.}~\bibnamefont {Padrick}}, \bibinfo {author} {\bibfnamefont
  {A.}~\bibnamefont {Mittal}}, \bibinfo {author} {\bibfnamefont
  {R.}~\bibnamefont {Ali}}, \bibinfo {author} {\bibfnamefont {A.}~\bibnamefont
  {Yunus}}, \bibinfo {author} {\bibfnamefont {D.}~\bibnamefont {Liu}}, \bibinfo
  {author} {\bibfnamefont {R.}~\bibnamefont {Pappu}},\ and\ \bibinfo {author}
  {\bibfnamefont {M.}~\bibnamefont {Rosen}},\ }\bibfield  {title} {\bibinfo
  {title} {{Sequence Determinants of Intracellular Phase Separation by Complex
  Coacervation of a Disordered Protein}},\ }\href
  {https://doi.org/10.1016/j.molcel.2016.05.042} {\bibfield  {journal}
  {\bibinfo  {journal} {Molecular Cell}\ }\textbf {\bibinfo {volume} {63}},\
  \bibinfo {pages} {72} (\bibinfo {year} {2016})}\BibitemShut {NoStop}%
\bibitem [{\citenamefont {Yin}\ \emph {et~al.}(2016)\citenamefont {Yin},
  \citenamefont {Niu}, \citenamefont {Zhu}, \citenamefont {Zhao}, \citenamefont
  {Zhang}, \citenamefont {Mann},\ and\ \citenamefont
  {Liang}}]{Yin2016Non-equilibriumExcitation}%
  \BibitemOpen
  \bibfield  {author} {\bibinfo {author} {\bibfnamefont {Y.}~\bibnamefont
  {Yin}}, \bibinfo {author} {\bibfnamefont {L.}~\bibnamefont {Niu}}, \bibinfo
  {author} {\bibfnamefont {X.}~\bibnamefont {Zhu}}, \bibinfo {author}
  {\bibfnamefont {M.}~\bibnamefont {Zhao}}, \bibinfo {author} {\bibfnamefont
  {Z.}~\bibnamefont {Zhang}}, \bibinfo {author} {\bibfnamefont
  {S.}~\bibnamefont {Mann}},\ and\ \bibinfo {author} {\bibfnamefont
  {D.}~\bibnamefont {Liang}},\ }\bibfield  {title} {\bibinfo {title}
  {{Non-equilibrium behaviour in coacervate-based protocells under
  electric-field-induced excitation}},\ }\href
  {https://doi.org/10.1038/ncomms10658} {\bibfield  {journal} {\bibinfo
  {journal} {Nature Communications}\ }\textbf {\bibinfo {volume} {7}},\
  \bibinfo {pages} {10658} (\bibinfo {year} {2016})}\BibitemShut {NoStop}%
\bibitem [{\citenamefont {Agrawal}\ \emph {et~al.}(2022)\citenamefont
  {Agrawal}, \citenamefont {Douglas}, \citenamefont {Tirrell},\ and\
  \citenamefont {Karim}}]{Agrawal2022ManipulationField}%
  \BibitemOpen
  \bibfield  {author} {\bibinfo {author} {\bibfnamefont {A.}~\bibnamefont
  {Agrawal}}, \bibinfo {author} {\bibfnamefont {J.~F.}\ \bibnamefont
  {Douglas}}, \bibinfo {author} {\bibfnamefont {M.}~\bibnamefont {Tirrell}},\
  and\ \bibinfo {author} {\bibfnamefont {A.}~\bibnamefont {Karim}},\ }\bibfield
   {title} {\bibinfo {title} {{Manipulation of coacervate droplets with an
  electric field}},\ }\href {https://doi.org/10.1073/pnas.2203483119}
  {\bibfield  {journal} {\bibinfo  {journal} {Proceedings of the National
  Academy of Sciences}\ }\textbf {\bibinfo {volume} {119}},\ \bibinfo {pages}
  {e2203483119} (\bibinfo {year} {2022})}\BibitemShut {NoStop}%
\bibitem [{\citenamefont {Chen}\ \emph {et~al.}(2022)\citenamefont {Chen},
  \citenamefont {Zhang},\ and\ \citenamefont
  {Wang}}]{Chen2022ComplexationAsymmetry}%
  \BibitemOpen
  \bibfield  {author} {\bibinfo {author} {\bibfnamefont {S.}~\bibnamefont
  {Chen}}, \bibinfo {author} {\bibfnamefont {P.}~\bibnamefont {Zhang}},\ and\
  \bibinfo {author} {\bibfnamefont {Z.-G.}\ \bibnamefont {Wang}},\ }\bibfield
  {title} {\bibinfo {title} {{Complexation between Oppositely Charged
  Polyelectrolytes in Dilute Solution: Effects of Charge Asymmetry}},\ }\href
  {https://doi.org/10.1021/acs.macromol.2c00339} {\bibfield  {journal}
  {\bibinfo  {journal} {Macromolecules}\ }\textbf {\bibinfo {volume} {55}},\
  \bibinfo {pages} {3898} (\bibinfo {year} {2022})}\BibitemShut {NoStop}%
\bibitem [{\citenamefont {Crabtree}\ \emph {et~al.}(2021)\citenamefont
  {Crabtree}, \citenamefont {Holland}, \citenamefont {Kompella}, \citenamefont
  {Babl}, \citenamefont {Turner}, \citenamefont {Baldwin},\ and\ \citenamefont
  {Nott}}]{Crabtree2021RepulsiveCondensates}%
  \BibitemOpen
  \bibfield  {author} {\bibinfo {author} {\bibfnamefont {M.~D.}\ \bibnamefont
  {Crabtree}}, \bibinfo {author} {\bibfnamefont {J.}~\bibnamefont {Holland}},
  \bibinfo {author} {\bibfnamefont {P.}~\bibnamefont {Kompella}}, \bibinfo
  {author} {\bibfnamefont {L.}~\bibnamefont {Babl}}, \bibinfo {author}
  {\bibfnamefont {N.}~\bibnamefont {Turner}}, \bibinfo {author} {\bibfnamefont
  {A.~J.}\ \bibnamefont {Baldwin}},\ and\ \bibinfo {author} {\bibfnamefont
  {T.~J.}\ \bibnamefont {Nott}},\ }\bibfield  {title} {\bibinfo {title}
  {{Repulsive electrostatic interactions modulate dense and dilute phase
  properties of biomolecular condensates}},\ }\href
  {https://doi.org/10.1101/2020.10.29.357863} {\bibfield  {journal} {\bibinfo
  {journal} {bioRxiv}\ ,\ \bibinfo {pages} {2020.10.29.357863}} (\bibinfo
  {year} {2021})}\BibitemShut {NoStop}%
\bibitem [{\citenamefont {Zhang}\ and\ \citenamefont
  {Shklovskii}(2005)}]{Zhang2005PhasePolyelectrolytes}%
  \BibitemOpen
  \bibfield  {author} {\bibinfo {author} {\bibfnamefont {R.}~\bibnamefont
  {Zhang}}\ and\ \bibinfo {author} {\bibfnamefont {B.}~\bibnamefont
  {Shklovskii}},\ }\bibfield  {title} {\bibinfo {title} {{Phase diagram of
  solution of oppositely charged polyelectrolytes}},\ }\href
  {https://doi.org/10.1016/j.physa.2004.12.037} {\bibfield  {journal} {\bibinfo
   {journal} {Physica A: Statistical Mechanics and its Applications}\ }\textbf
  {\bibinfo {volume} {352}},\ \bibinfo {pages} {216} (\bibinfo {year}
  {2005})}\BibitemShut {NoStop}%
\bibitem [{\citenamefont {Groot}(2003)}]{Groot2003ElectrostaticSurfactants}%
  \BibitemOpen
  \bibfield  {author} {\bibinfo {author} {\bibfnamefont {R.~D.}\ \bibnamefont
  {Groot}},\ }\bibfield  {title} {\bibinfo {title} {{Electrostatic interactions
  in dissipative particle dynamics—simulation of polyelectrolytes and anionic
  surfactants}},\ }\href {https://doi.org/10.1063/1.1574800} {\bibfield
  {journal} {\bibinfo  {journal} {The Journal of Chemical Physics}\ }\textbf
  {\bibinfo {volume} {118}},\ \bibinfo {pages} {11265} (\bibinfo {year}
  {2003})}\BibitemShut {NoStop}%
\bibitem [{\citenamefont {Groot}\ and\ \citenamefont
  {Warren}(1997)}]{Groot1997DissipativeSimulation}%
  \BibitemOpen
  \bibfield  {author} {\bibinfo {author} {\bibfnamefont {R.~D.}\ \bibnamefont
  {Groot}}\ and\ \bibinfo {author} {\bibfnamefont {P.~B.}\ \bibnamefont
  {Warren}},\ }\bibfield  {title} {\bibinfo {title} {{Dissipative particle
  dynamics: Bridging the gap between atomistic and mesoscopic simulation}},\
  }\href {https://doi.org/10.1063/1.474784} {\bibfield  {journal} {\bibinfo
  {journal} {The Journal of Chemical Physics}\ }\textbf {\bibinfo {volume}
  {107}},\ \bibinfo {pages} {4423} (\bibinfo {year} {1997})}\BibitemShut
  {NoStop}%
\bibitem [{\citenamefont {Delaney}\ and\ \citenamefont
  {Fredrickson}(2017)}]{Delaney2017TheorySelf-coacervates}%
  \BibitemOpen
  \bibfield  {author} {\bibinfo {author} {\bibfnamefont {K.~T.}\ \bibnamefont
  {Delaney}}\ and\ \bibinfo {author} {\bibfnamefont {G.~H.}\ \bibnamefont
  {Fredrickson}},\ }\bibfield  {title} {\bibinfo {title} {{Theory of
  polyelectrolyte complexation—Complex coacervates are self-coacervates}},\
  }\href {https://doi.org/10.1063/1.4985568} {\bibfield  {journal} {\bibinfo
  {journal} {The Journal of Chemical Physics}\ }\textbf {\bibinfo {volume}
  {146}},\ \bibinfo {pages} {224902} (\bibinfo {year} {2017})}\BibitemShut
  {NoStop}%
\bibitem [{\citenamefont {Hutchens}\ and\ \citenamefont
  {Wang}(2007)}]{Hutchens2007MetastableSolutions}%
  \BibitemOpen
  \bibfield  {author} {\bibinfo {author} {\bibfnamefont {S.~B.}\ \bibnamefont
  {Hutchens}}\ and\ \bibinfo {author} {\bibfnamefont {Z.-G.}\ \bibnamefont
  {Wang}},\ }\bibfield  {title} {\bibinfo {title} {{Metastable cluster
  intermediates in the condensation of charged macromolecule solutions}},\
  }\href {https://doi.org/10.1063/1.2761891} {\bibfield  {journal} {\bibinfo
  {journal} {The Journal of Chemical Physics}\ }\textbf {\bibinfo {volume}
  {127}},\ \bibinfo {pages} {084912} (\bibinfo {year} {2007})}\BibitemShut
  {NoStop}%
\bibitem [{\citenamefont {Welsh}\ \emph {et~al.}(2022)\citenamefont {Welsh},
  \citenamefont {Krainer}, \citenamefont {Espinosa}, \citenamefont {Joseph},
  \citenamefont {Sridhar}, \citenamefont {Jahnel}, \citenamefont {Arter},
  \citenamefont {Saar}, \citenamefont {Alberti}, \citenamefont
  {Collepardo-Guevara},\ and\ \citenamefont
  {Knowles}}]{Welsh2022SurfaceCondensates}%
  \BibitemOpen
  \bibfield  {author} {\bibinfo {author} {\bibfnamefont {T.~J.}\ \bibnamefont
  {Welsh}}, \bibinfo {author} {\bibfnamefont {G.}~\bibnamefont {Krainer}},
  \bibinfo {author} {\bibfnamefont {J.~R.}\ \bibnamefont {Espinosa}}, \bibinfo
  {author} {\bibfnamefont {J.~A.}\ \bibnamefont {Joseph}}, \bibinfo {author}
  {\bibfnamefont {A.}~\bibnamefont {Sridhar}}, \bibinfo {author} {\bibfnamefont
  {M.}~\bibnamefont {Jahnel}}, \bibinfo {author} {\bibfnamefont {W.~E.}\
  \bibnamefont {Arter}}, \bibinfo {author} {\bibfnamefont {K.~L.}\ \bibnamefont
  {Saar}}, \bibinfo {author} {\bibfnamefont {S.}~\bibnamefont {Alberti}},
  \bibinfo {author} {\bibfnamefont {R.}~\bibnamefont {Collepardo-Guevara}},\
  and\ \bibinfo {author} {\bibfnamefont {T.~P.~J.}\ \bibnamefont {Knowles}},\
  }\bibfield  {title} {\bibinfo {title} {{Surface Electrostatics Govern the
  Emulsion Stability of Biomolecular Condensates}},\ }\href
  {https://doi.org/10.1021/acs.nanolett.1c03138} {\bibfield  {journal}
  {\bibinfo  {journal} {Nano Letters}\ }\textbf {\bibinfo {volume} {22}},\
  \bibinfo {pages} {612} (\bibinfo {year} {2022})}\BibitemShut {NoStop}%
\bibitem [{\citenamefont {Dai}\ \emph {et~al.}(2023)\citenamefont {Dai},
  \citenamefont {Chamberlayne}, \citenamefont {Messina}, \citenamefont {Chang},
  \citenamefont {Zare}, \citenamefont {You},\ and\ \citenamefont
  {Chilkoti}}]{Dai2023InterfaceReactions}%
  \BibitemOpen
  \bibfield  {author} {\bibinfo {author} {\bibfnamefont {Y.}~\bibnamefont
  {Dai}}, \bibinfo {author} {\bibfnamefont {C.~F.}\ \bibnamefont
  {Chamberlayne}}, \bibinfo {author} {\bibfnamefont {M.~S.}\ \bibnamefont
  {Messina}}, \bibinfo {author} {\bibfnamefont {C.~J.}\ \bibnamefont {Chang}},
  \bibinfo {author} {\bibfnamefont {R.~N.}\ \bibnamefont {Zare}}, \bibinfo
  {author} {\bibfnamefont {L.}~\bibnamefont {You}},\ and\ \bibinfo {author}
  {\bibfnamefont {A.}~\bibnamefont {Chilkoti}},\ }\bibfield  {title} {\bibinfo
  {title} {{Interface of biomolecular condensates modulates redox reactions}},\
  }\href {https://doi.org/10.1016/j.chempr.2023.04.001} {\bibfield  {journal}
  {\bibinfo  {journal} {Chem}\ }\textbf {\bibinfo {volume} {9}},\ \bibinfo
  {pages} {1594} (\bibinfo {year} {2023})}\BibitemShut {NoStop}%
\bibitem [{\citenamefont {Hu}\ \emph {et~al.}(2021)\citenamefont {Hu},
  \citenamefont {Relton}, \citenamefont {Locker}, \citenamefont {Phan},\ and\
  \citenamefont {Ewing}}]{Hu2021ElectrochemicalGranules}%
  \BibitemOpen
  \bibfield  {author} {\bibinfo {author} {\bibfnamefont {K.}~\bibnamefont
  {Hu}}, \bibinfo {author} {\bibfnamefont {E.}~\bibnamefont {Relton}}, \bibinfo
  {author} {\bibfnamefont {N.}~\bibnamefont {Locker}}, \bibinfo {author}
  {\bibfnamefont {N.~T.~N.}\ \bibnamefont {Phan}},\ and\ \bibinfo {author}
  {\bibfnamefont {A.~G.}\ \bibnamefont {Ewing}},\ }\bibfield  {title} {\bibinfo
  {title} {{Electrochemical Measurements Reveal Reactive Oxygen Species in
  Stress Granules}},\ }\href {https://doi.org/10.1002/anie.202104308}
  {\bibfield  {journal} {\bibinfo  {journal} {Angewandte Chemie International
  Edition}\ }\textbf {\bibinfo {volume} {60}},\ \bibinfo {pages} {15302}
  (\bibinfo {year} {2021})}\BibitemShut {NoStop}%
\end{thebibliography}%

\end{document}